\documentstyle[12pt]{article}
\advance \topmargin by -\headheight
\advance \topmargin by -\headsep
     
\evensidemargin \oddsidemargin
\begin{document}
\topmargin -.6in
\def\br{\begin{eqnarray}}
\def\er{\end{eqnarray}}
\def\be{\begin{equation}}
\def\ee{\end{equation}}
\def\({\left(}
\def\){\right)}
\def\a{\alpha}
\def\b{\beta}
\def\d{\delta}
\def\D{\Delta}
\def\g{\gamma}
\def\G{\Gamma}
\def\h{ {1\over 2}  }
\def\hp{ {+{1\over 2}}  }
\def\hm{ {-{1\over 2}}  }
\def\k{\kappa}
\def\l{\lambda}
\def\L{\Lambda}
\def\m{\mu}
\def\n{\nu}
\def\o{\over}
\def\O{\Omega}
\def\p{\phi}
\def\rh{\rho}
\def\s{\sigma}
\def\t{\tau}
\def\th{\theta}
\def\ii {\'\i  }

\begin{center}
{\large {\bf Induced Variational Method from Supersymmetric Quantum Mechanics and the Screened
Coulomb Potential}} \footnotemark
\end{center}
\normalsize
\vskip 1cm
\begin{center}

{\it  Elso Drigo Filho $^a$ \footnotemark
\footnotetext{Work supported in part by CNPq} 
and Regina Maria  Ricotta $^b$} \\
$^a$ Instituto de Bioci\^encias, Letras e Ci\^encias Exatas, IBILCE-UNESP\\
Rua Cristov\~ao Colombo, 2265 -  15054-000 S\~ao Jos\'e do Rio Preto - SP\\
$^b$  Faculdade de Tecnologia de S\~ao Paulo, FATEC/SP-CEETPS-UNESP  \\
Pra\c ca  Fernando Prestes, 30 -  01124-060 S\~ao Paulo-SP\\ 
Brazil\\
\end{center}
\vskip 2cm
{\bf  Abstract}\\
The formalism  of \-Supersymmetric Quantum Mechanics supplies a trial wave function to be used
in the Variational Method. The screened Coulomb potential is analysed within this approach.
Numerical and exact  results for energy eigenvalues are compared.
\\

\noindent   {\bf I. Introduction}\\

The screened Coulomb potential has been used in several branches of Physics,
for instance, in nuclear physics (as the name of Yukawa potential), in
plasma and in the study of electrolytic solution properties (Debye-Huckel potential).   The
Schr$\ddot {o}$dinger equation for this potential is not exactly solvable
and  exact numerical,
\cite{Rogers}, \cite{Roussel} and approximative, \cite{Imbo}-\cite{Dutt} methods have been
applied to obtain the energy eigenvalues, including variational
calculations, \cite{Greene}-\cite{Fessatidis}.

More recently, a new methodology based within  the variational method  associated to
supersymmetric quantum mechanics formalism has been introduced, \cite{Gozzi}-\cite{Drigo}.
References \cite{Gozzi}  and  \cite{Cooper1} introduce a scheme based in the hierarchy of
Hamiltonians; it permits the evaluation of excited states for one-dimensional systems. 
In reference \cite {Drigo}  an {\it ansatz} for the superpotential which is  related
to the trial wave function is proposed. The new approach showed to be useful to get answers  when 
applied to atomic systems, \cite {Drigo}-\cite {Drigo1}.

In this letter, the variational energy  eigenvalues for the static screened
Coulomb potential in three dimensions are determined using the variational
method using a trial wave function  induced by Supersymmetric Quantum
Mechanics, SQM.

In the approach followed here the first step taken is to look for an
effective potential similar to the original screened Coulomb potential.
Inspired by SQM, an {\it  ansatz} is made to the superpotential which
determines the variational (trial)  wavefunction through the algebraic
approach of SQM.

Our system is three dimensional and in this case it is possible to determine
the variational  eigenfunctions for each value of angular momentum $\it l$.
The first eigenfunction, obtained from direct factorization of the effective
Hamiltonian, corresponds to the minimum energy for each 
$\it l$.

This new methodology has been successfully applied to other atomic systems such as the
Hulth\'en, \cite{Drigo},  and the Morse, \cite{Drigo1}, potentials.  Here it is applied to the
screened Coulomb potential.

In what follows, we briefly introduce, for completeness,  the SQM scheme, 
then introduce  the variational method and show our results. \\

\noindent {\bf  II. Supersymmetric Quantum Mechanics}\\

In SQM, \cite{Cooper1}-\cite{Drigo}, for $N=2$ we have two
nilpotent operators, $Q$ and $Q^+$, that satisfying the algebra
\be
\{ Q, Q^+\} = H_{SS} \;\;\;\;; Q^2  = {Q^+}^2 = 0, 
\ee
where $H_{SS}$ is the supersymmetric Hamiltonian. This algebra can be realized as
\be
Q =  \left( \begin{array}{cc} 0  & 0  \\ A^-  & 0 
\end{array} \right ) \;,\;\;\;
Q^+ = \left( \begin{array}{cc} 0  & A^+  \\ 0 & 0 
\end{array} \right )
\ee
where $A^{\pm}$ are bosonic operators.  With this realization the supersymmetric Hamiltonian 
$H_{SS}$ is then given by
\be
H_{SS} = \left( \begin{array}{cc} A^+A^-  &  0 \\ 0 & A^-A^+
\end{array} \right ) = \left( \begin{array}{cc} H^+  &  0 \\ 0 & H^-
\end{array} \right ).
\ee
where $H^{\pm}$ are called supersymmetric partner Hamiltonians and  share the same spectra, apart
from the nondegenerate ground state.  Using the super-algebra, a given Hamiltonian can be
factorized in terms of the bosonic operators. In $\hbar = c = 1$ units, it is given by
\be
H_1 =  -{1\o 2}{d^2 \o d r^2} + V_1(r) =  A_1^+A_1^-  + E_0^{(1)} 
\ee
where $ E_0^{(1)}$ is the lowest eigenvalue and the function $V_1(r)$ includes the barrier
potential term.  The bosonic operators are  defined by 
\be 
A_1^{\pm} =  {1\o \sqrt 2}\left(\mp {d \o dr} + W_1(r) \right) 
\ee
where the superpotential $W_1(r)$ satisfies the Riccati equation
\be
\label{Riccati}
W_1^2 - W_1'=  2V_1(r) - 2E_0^{(1)} 
\ee   
which is a consequence of the factorization of the Hamiltonian $H_1$.

The eigenfunction for the lowest state is related to the superpotential $W_1$ by
\be
\label{eigenfunction}
\Psi_0^{(1)} (r) = N exp( -\int_0^r W_1(\bar r) d\bar r).
\ee
Now it is possible to construct the supersymmetric partner Hamiltonian
\be
H_2 = A_1^-A_1^+ + E_0^{(1)} =  -{1\o 2}{d^2 \o d r^2} + {1\o 2}(W_1^2 +
W_1')+ E_0^{(1)} .
\ee
If one factorizes  $H_2$ in terms of a new pair of bosonic operators,
$A_2^{\pm}$ one gets,
\be
H_2 = A_2^+A_2^- + E_0^{(2)} =  -{1\o 2}{d^2 \o d r^2} + {1\o 2}(W_2^2 -
W_2')+ E_0^{(2)} 
\ee
where $E_0^{(2)} $ is the lowest eigenvalue of $H_2$ and $W_2$ satisfy the
Riccati equation,
\be
W_2^2 - W_2'=  2V_2(r) - 2E_0^{(2)}  .
\ee
Thus a whole hierarchy of Hamiltonians can be constructed, with simple
relations connecting the eigenvalues and eigenfunctions of the $n$-members, 
\cite{Sukumar1}, \cite{Cooper2}.

Thus, the formalism of  SQM allows us to evaluate the ground state
eigenfunction from the knowledge of the superpotential  $W(r)$, satisfying the  Riccati equation,
eq.(\ref {Riccati}).  However, since the  potential is not exactly solvable, the purpose is to
propose an {\it ansatz} for the  superpotential and, based  in the
superalgebra information, we evaluate a trial wavefunction that minimizes the
expectation  value of the energy. The energy eigenvalues pursued are
evaluated by minimizing the energy expectation value with respect to a free
parameter introduced by the {\it ansatz}. \\

\noindent {\bf  III. Trial wavefunction for the variational calculation}\\

The screened Coulomb potential is given, in atomic units, by 
\be
\label{Coulomb}
V_{SC} = - {e^{-\d r} \o r}
\ee
where $\delta$ is the screened length. The associated radial Schr$\ddot
{o}$dinger equation includes the potential barrier term and it is given by
\be
\label{Hamiltonian}
 \left(-{1\o 2}{d^2 \o d r^2} - {e^{-\d r} \o r} + { l(l + 1) \o
2r^2}\right) \Psi  = E \Psi
\ee
where the unit length is $\hbar^2 /me^2 $ and the energy unit is
$\epsilon_0 = - me^4/\hbar^2$. 

In order to determine an effective potential similar to the potential in
the Hamiltonian (\ref{Hamiltonian}), that is the screened Coulomb potential
plus the potential barrier term, the following {\it ansatz} to the superpotential is suggested 
\be
\label{superpotential}
W(r) = - (l+1){\d e^{-\d r} \o 1 - e^{-\d r} }+ {1\o  (l+1)} - {\d
\o 2}.
\ee
Substituting it into (\ref{eigenfunction}), one gets
\be
\label{Psi}
\Psi_0 (r) = (1-e^{-\d r})^{l+1} e^{- ({1\o  (l+1)} - {\d\o 2})r}.
\ee
Assuming that the radial trial wave function is given by (\ref{Psi}),
replacing $\d $ by the variational parameter $\mu $, i.e., 
\be
\label{Psimu}
\Psi_{\mu} (r) = (1-e^{-\mu r})^{l+1} e^{ -({1\o  (l+1)} - {\mu \o 2})r},
\ee
the variational energy is given by
\be
\label{energymu}
E_{\mu} = {\int_0^{\infty} \Psi_{\mu}(r) [\hm {d^2 \o dr^2} -               
{ e^{-\d r}\o r} + {l(l+1)\o 2r^2}] \Psi_{\mu}(r) dr
\o \int_0^{\infty} \Psi_{\mu}(r)^2 dr}.
\ee
Thus, minimizing this energy with respect to the variational parameter $\mu$ one
obtains the best estimate for the energy of the screened Coulomb potential.

As our potential is not exactly solvable, the  superpotential given by eq.(\ref{superpotential}) 
does not satisfy the Riccati equation (\ref{Riccati}) but it does satisfy it for an effective
potential instead,
$V_{eff}$ 
\be
V_{eff}(r) = {\bar W_1^2 - \bar W_1' \o 2}+ E(\bar\mu)    
\ee
where $ \bar W_1 = W_1(\d=\bar\mu)$ is given by  eq.(\ref{superpotential}) and 
$\bar\mu$ is the parameter that minimises the energy expectation value, (\ref{energymu}). It is
given by
\be
\label{effective}
V_{eff}(r) = - {\d e^{-\d r}\o 1-e^{-\d r}} + {l(l+1)\o 2}{{\d}^2  e^{-2\d
r}\o (1-e^{-\d r})^2} + {1\o 2}({1\o l+1} - {\d\o 2})^2  + E(\d),
\ee
where $\d = \bar\mu$ that minimises energy expectation value. One observes
that for small values of 
$\d$ the first term is similar to the potential (\ref{Coulomb}) and the last is 
approximately the potential barrier term.  This observation allows us to
conclude that the superpotential (\ref{superpotential})  can be used to
analyse the three dimensional screened Coulomb potential variationally through the trial
wavefunction (\ref{Psi}).\\

\noindent {\bf  IV. Results}\\

For $l=0$ the effective potential (\ref{effective}) becomes identical to
the Hulth\'en potential. Thus, the results presented in Table 1 coincide with those of 
ref.\cite{Lam}, where the Hulth\'en potential eigenfunctions are directly used in the variational
calculation.  The deviation on  the fifth decimal algarism can be attributed to the accuracy of
the numerical calculation.

\begin{tabular}{|c|c|c|c|} \hline
\multicolumn{4}{ |c|} 
{State 1s}\\ \hline
{Screening $\delta$} & SQM Variational & Variational (Ref. 7) & Exact Numerical \\ \hline
0.001 & -0.49902  & -0.49900  & -\\ \hline
0.002 & 0.49802 & -0.49800 & -0.4980\\ \hline
0.005 & -0.49504  & -0.49502 & -0.4950\\ \hline 
0.010 & -0.49009 & -0.49007 & -0.4901 \\ \hline 
0.02 & -0.48031 & -0.48030 & -0.4803\\ \hline
0.025  & -047548 & -0.47546 & -0.4755\\ \hline
0.03 & -0.47068 & -0.47066 & -\\ \hline
0.04 & -0.46119  &-0.46117  & -\\ \hline
0.05& -0.45180 &-0.45182 & -0.4518\\ \hline 
0.06 & -0.44259  & -0.44260 & -\\ \hline 
0.07 & -0.43351 & -0.43352 & -\\ \hline
0.08 & -0.42456 & -0.42457 & -\\ \hline 
0.09 & -0.41574 & -0.41575  & -\\ \hline
0.10 & -0.40705 & -0.40706  & -0.4071\\ \hline
0.20 & -0.32681 & -0.32681 & -0.3268\\ \hline
0.25 & -0.29092 & -0.29092 &  -0.2909\\ \hline
0.30 &  -0.25764& -0.25763 & -\\ \hline
0.40 & -0.19842 & -0.19836 &  -\\ \hline
0.50 & -0.14806 & -0.14808 & -0.1481\\ \hline
0.60 & -0106077 & -0.10608 & -\\ \hline
0.70 & -0.07175 & -0.07174 & -\\ \hline
0.80 & -0.04459 & -0.04459 & -\\ \hline
0.90 & -0.02420 & -0.02418 & -\\ \hline
1.00 & -0.01026 & -0.01016 & -0.01029\\ \hline
1.05 & -0.00568 & -0.00544 & -\\ \hline
\end{tabular}\\
\vskip 1cm
{\bf Table 1.} Energy eigenvalues as function of the  screening
parameters $\d$ for $1 s$ state ($l=0$),  in rydberg units of energy.  Comparison is make with
variational and exact numerical results from \cite{Rogers} and \cite{Lam}.\\

The results become more interesting for $l\not=0$. In this case the 
effective potential differs from the Hulth\'en potential. Table 2 shows the results for $2p$ ($l=1$), $3d $ ($l=2$) and $4f$ ($l=3$)
energy levels.  Also given in this table are the correponding numerical results \cite{Rogers}. \\
 \\

\vskip .5cm
\begin{tabular}{|c|c|c|c|c|c|c|} \hline
\multicolumn{1}{|c} {} &
\multicolumn{2}{|c} {2p }   &
\multicolumn{2}{|c|} {3d } & 
\multicolumn{2}{|c|} {4f }  \\ \hline 
{$ \;\;\delta \;\;$} 
& {Variational} & {Numerical}   &
{Variational} & {Numerical} & 
{Variational} & {Numerical}  \\ \hline 
0.001 & -0.2480 & -0.2480 & -0.10910 &-0.10910  & -0.06051 & -0.06052 \\
\hline    0.005 &  - &  - &  - &  - & -0.52930 & -0.05294 \\ \hline  
0.010 & -0.2305 & -0.2305  & -0.09212 & -0.09212   &  -0.04419 &  -0.04420 \\ \hline  
0.020 & -0.2119 & -0.2119 & -0.07503 & -0.07503 & -0.02897 & -0.02898 \\ \hline  
0.025 & -0.2030 & -0.2030  & -0.06714 & -0.06715 & - & - \\ \hline  
0.050 & -0.1615 & -0.1615  & -0.03374 & -0.03383 & - &  - \\ \hline 
0.100 & -0.09289 & -0.09307   &  - &  - &   - &  - \\ \hline
\end{tabular}\\
\vskip 1cm
{\bf Table 2.} Energy eigenvalues as function of the  screening
parameters $\d$ for $2p$ ($l=1$), $3d $ ($l=2$) and $4f$ ($l=3$) states, in rydberg units of
energy. Variational values obtained by the trial function (\ref{Psimu})  are compared with exact  numerical  results obtained
from reference \cite{Rogers}, (see also \cite{Greene}).\\

{\bf V. Conclusions}\\

We have proposed trial wavefunctions to be used in the variational calculation
in order to determine the energy eigenvalues of the screened Coulomb
potential.  These functions were induced from supersymmetric quantum
mechanics formalism.  The approach consists of making an {\it ansatz} in the superpotential
which satifies the Riccati equation by an effective potential.  The trial wavefunctions were then
determined from this superpotential through the superalgebra. 

For $l=0$ the effective potential obtained is identical to the Hulth\'en
potential.  However for $l\not=0$ the potential has a new structure.  The
trial wavefunctions suggested for this case are different from those proposed
in references \cite{Greene}-\cite{Fessatidis}. Within our framework the energy  eigenvalue for
each value of $l$ is obtained using the same function (\ref{Psi}).

In terms of the hierarchy of Hamiltonians, we obtained the first member for
each value of $l$.  Other members can be determined from the usual approach
in supersymmetric quantum mechanics, \cite{Cooper2}.

One observes that the results obtained are in very good agreement to those found in the
literature.  The results are better for small values of the parameter $\d$. 
This observation is justified by the fact that for small values of $\d $ the
effective potential is more similar to the original potential than for
higher values of $\d$.  

We stress that even though the problem has been attacked by different methods our new
methodology is very simple to supply accurate results.  We believe that other applications to
atomic physics problems can be made by this new method.\\

\end{document}